\documentclass[twocolumn,superscriptaddress,aps,preprintnumbers,amsmath,amssymb,prl,nofootinbib]{revtex4}

\usepackage{amsmath}
\usepackage{amssymb}
\usepackage{amsfonts}
\usepackage{graphicx}
\usepackage{xcolor}
\usepackage{xfrac}
\usepackage{comment}
\usepackage{pifont}
\usepackage{times}
\usepackage{hyperref}
\usepackage{bm}
\usepackage{enumitem}
\usepackage{centernot}
\usepackage{cancel}

\usepackage{amsmath,amssymb,mathrsfs}
\usepackage{bm}
\usepackage{times}
\usepackage{epsfig}
\usepackage{verbatim}
\usepackage{bm}
\usepackage[utf8]{inputenc}
\usepackage{graphics}
\usepackage{graphicx,epsfig,amssymb,amsmath,color,cancel}
\usepackage{subfigure}
\usepackage[english]{babel}

\begin{document}

\preprint{DESY 21-126}

\title{Revealing the Primordial Irreducible Inflationary Gravitational-Wave\\ Background with a Spinning Peccei-Quinn Axion}

\author{Yann Gouttenoire}
\affiliation{School of Physics and Astronomy, Tel-Aviv University, Tel-Aviv 69978, Israel}
\author{G\'eraldine Servant}
\affiliation{DESY, Notkestra{\ss}e 85, D-22607 Hamburg, Germany}
\affiliation{II. Institute of Theoretical Physics, University of Hamburg, D-22761 Hamburg, Germany}
\author{Peera Simakachorn}
\affiliation{DESY, Notkestra{\ss}e 85, D-22607 Hamburg, Germany}
\affiliation{II. Institute of Theoretical Physics, University of Hamburg, D-22761 Hamburg, Germany}

\date{\today}

\begin{abstract}

 The primordial irreducible  gravitational-wave background 
 due to quantum vacuum tensor fluctuations produced during inflation spans a large range of frequencies with an almost scale-invariant spectrum but is too low to be detected by the next generation of gravitational-wave interferometers. 
We show how this signal is enhanced by a short  temporary kination era in the cosmological history (less than 10 e-folds), that can arise at any energy scale between a GeV and the inflationary scale $10^{16}$ GeV.
We argue that such kination era is naturally generated  by a spinning axion before it gets trapped by its potential.
It is usually assumed that the axion starts oscillating around its minimum from its initially frozen position.
However, the early dynamics of the Peccei-Quinn field can induce a large kinetic energy in the axion field, triggering a kination era, either before or after the axion acquires its mass, leading to a characteristic peak in the primordial gravitational-wave background. This represents a smoking-gun signature of axion physics as no other scalar field dynamics is expected to trigger such a sequence of equations of state in the early universe.
We derive the resulting gravitational-wave spectrum, and present the parameter space that leads to such a signal as well as the detectability prospects, in particular at LISA, Einstein Telescope, Cosmic Explorer and Big Bang Observer.
We show both model-independent predictions and present as well results for two specific well-motivated UV completions for the QCD axion dark matter where this dynamics is built-in.

\end{abstract}

\maketitle

\section{Introduction}

Axion particles are ubiquitous in extensions of the Standard Model of particle physics.
They arise as pseudo-Nambu-Goldstone bosons of a spontaneously broken global $U(1)$ symmetry and as such are typically very light compared to the symmetry-breaking energy scale $f_a$. 
A particularly well-motivated candidate is the QCD axion predicted by the Peccei-Quinn (PQ) mechanism introduced to solve the strong CP problem in the Standard Model, which is the intriguing absence of CP-violation in the sector of strong interactions. 
The PQ mechanism relies on the existence of a new complex scalar field whose vacuum expectation value  breaks spontaneously a new broken global $U(1)_{PQ}$ symmetry.
Such mechanism still remains to be tested experimentally.
Its main prediction is  a new light particle, the axion, the angular mode of the new PQ scalar field.
There has been growing interest for the axion over the years, as it can as well explain the Dark Matter of the Universe.
The axion is at the origin of an extensive experimental programme, and has become  the most hunted particle after the Higgs discovery. 
It is being searched by exploiting its coupling to the photon which scales as $1/ f_a$. Given the astrophysical constraints on $f_a\gtrsim 10^8$ GeV, the small axion coupling makes its detection challenging.
Around the QCD epoch, the QCD axion acquires a tiny mass, $m_a \propto \Lambda_{QCD}^2/f_a$. From that time, it starts oscillating around its minimum, and its energy density redshifts as pressure-less matter $a^{-3}$.
It is usually assumed that the axion starts oscillating from its initially frozen position.
However, the early dynamics for the Peccei-Quinn field before the QCD scale can naturally induce a large kinetic energy  for the axion, which thus may experience a fast-rotating stage before oscillations kick in. This is the scenario we investigate in this work, establishing a gravitational-wave smoking-gun signature of this dynamics.

\section{Gravitational-Waves from a kination era}


Today, the irreducible stochastic background of gravitational waves from inflation, characterized by its cosmological fraction of the total energy density, reads \cite{Caprini:2018mtu}
\begin{equation}
\Omega_{\rm GW}= \frac{k^2a_k^2}{24H_0^2} \Omega_{\rm GW,inf}
\label{eq:Omega}
\end{equation}
and comes from modes  with comoving wave number 
\begin{equation}
k=a_k H_k 
\end{equation}
  which entered the horizon when the scale factor of the universe was $a_k$
and the expansion rate of the universe was $H_k$. $H_0$ is the Hubble rate today.
GW generated during inflation are stretched outside the Hubble horizon and are well-known to represent a mainly scale-invariant power spectrum when they enter the horizon:
\begin{equation}
\Omega_{\rm GW,inf} \simeq \frac{2}{\pi^2} \left( \frac{H_{I}}{M_{\rm pl}} \right)^2,\label{eq:primordial_inf_spectrum}
\end{equation}
where $H_I$ is the Hubble rate during inflation.
The frequency of GW we observe today is
\begin{equation}
f =  \frac{H_k}{2 \pi} \frac{a_k }{a_0} 
\end{equation}
Using the Friedmann equation $H={\sqrt{\rho/3}} M_{\rm Pl}$
where $\rho\propto a^{-3(1+\omega)}$, $\omega$ being the equation of state of the universe, we have $f \propto a_k^{-{(1+3\omega)}/{2}}$ and Eq.~(\ref{eq:Omega}) leads to
\begin{equation}
\Omega_{\rm GW}
 \propto ~ f^{\beta}, ~ ~ \textrm{with} ~ ~ \beta ~ \equiv ~ -2\left(\frac{1-3\omega}{1+3\omega}\right),
 \label{eq:spectralindex}
\end{equation}
\begin{figure}[t]
\includegraphics[width=0.475\textwidth]{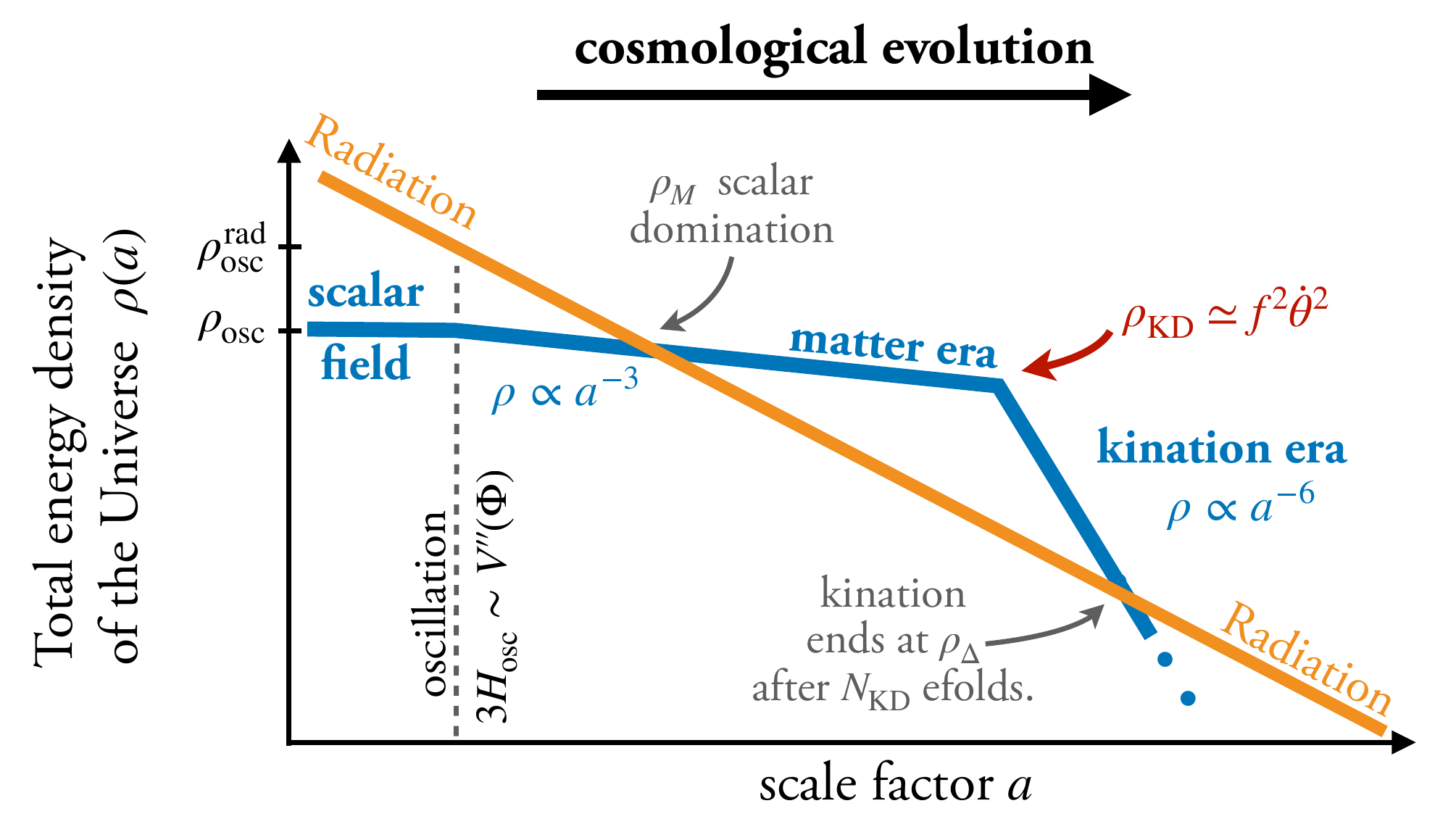}
\caption{\small \it{
Kination era generated by the dynamics of the Peccei-Quinn field. First, the energy density of the universe is dominated by the scalar field oscillations.   Kination starts once the kinetic energy of the scalar field dominates.}}
\label{fig:diag}
\end{figure}
Therefore, modes entering the horizon during radiation ($\omega=1/3$), matter ($\omega=0$) and kination ($\omega=1$)
eras have spectral indices $\beta = 0, -2$, and 1, respectively.
GW resulting from modes that enter during the radiation era have the standard flat spectrum 
\begin{equation}
\Omega^\mathrm{st}_{\rm GW} h^2  \simeq  (1.29  \times 10^{-17}) G(T_k)  \left(\frac{V_\mathrm{inf}^{1/4}}{10^{16}\textrm{ GeV}}\right)^4,
\label{st_GW_inflation}
\end{equation}
where $G(T_k)=  ({g_*(T_k)}/{106.75}) ({g_{*,s}(T_k)}/{106.75})^{-4/3}$, $T_k$ is the temperature when a given mode re-enters the horizon, and $V_{\rm inf}^{1/4}$ is the inflationary energy scale.
Therefore, even assuming the largest inflation energy scale allowed by CMB data \cite{Akrami:2018odb}, this GW background cannot be observed by the future GW observatories LISA \cite{Audley:2017drz} and Einstein Telescope \cite{Hild:2010id, Punturo:2010zz}. Only Big Bang Observer \cite{Yagi:2011wg} may be sensitive  to it.
In this letter, we show how axion models produce a kination era preceeded by a matter era inside the standard radiation era, as illustrated in Fig.~\ref{fig:diag}. 
The transition between these eras provides a sign change in the spectral index and leads to a peaked GW signature.  The high-frequency slope -2 is associated to the matter era
while the low-frequency slope  +1  is associated to the kination era.
The overall GW spectrum over frequency range can be written as
\begin{eqnarray}
\Omega_\mathrm{GW,0} h^2 (f) & &=  \Omega^\mathrm{st}_{\rm GW}(f_\Delta)h^2 \times \\
&&\begin{cases}
1 &;  f < f_\Delta,\\
\left(f/f_\Delta\right)  &; f_\Delta < f < f_\mathrm{KD},\\
\left(f_\mathrm{KD}/f_\Delta\right) \left(f_\mathrm{KD}/f\right)^2  &; f_\mathrm{KD} < f < f_{M},\\
\left(f_\mathrm{KD}/f_\Delta\right) \left(f_\mathrm{KD}/f_M\right)^2 &; f_{M} < f,
\end{cases}
\nonumber
\label{inflation_GW_master}
\end{eqnarray}
where the GW in standard cosmology $  \Omega^\mathrm{st}_{\rm GW}$ is given by Eq. (\ref{st_GW_inflation}), and
$f_\Delta$, $f_\mathrm{KD}$ (peak frequency), $f_M$ are the characteristic frequencies corresponding to the modes re-entering the horizon right after the end of the
kination era, at the beginning of the kination era, and at the beginning of the matter era respectively.
 They are defined as:\\
\begin{eqnarray}
f_\Delta  &&=  \frac{H_\Delta a_\Delta}{2 \pi a_0}   \simeq   2.6 \times 10^{-6}  \textrm{ Hz} \\
&&\times \left(\frac{g_*(T_{\Delta})}{106.75}\right)^{1/2} \left(\frac{g_{*,s}(T_{\Delta})}{106.75}\right)^{-1/3} \left(\frac{T_{\Delta}}{10^{2}\textrm{ GeV}}\right),
\nonumber
\end{eqnarray}
%
\begin{eqnarray}
\label{bump_peak_kination_inf}
f_\textrm{KD} &=&  \frac{H_\textrm{KD} a_\textrm{KD}}{2 \pi a_0}  
=  f_\Delta  \left(\frac{\rho_\textrm{KD}}{\rho_\Delta} \right)^{1/3}
 =  f_\Delta e^{2 N_\mathrm{KD}}\\
&\simeq & 1.07 \times 10^{-3}  \textrm{Hz} \times G^{1/4}(T_\Delta)
\times \left(\frac{\rho_\textrm{KD}^{1/4}}{\textrm{10 TeV}}\right) \frac{e^{N_\textrm{KD}/2}}{10}.
\nonumber
\end{eqnarray}
where the e-folding of the kination era is $e^{N_\textrm{KD}} = (\rho_\mathrm{KD}/\rho_\Delta)^{1/6}$.
This peak frequency $f_\textrm{KD} $ thus encodes information about the duration of the kination era.
The peak amplitude at $f_\mathrm{KD}$ is
\begin{eqnarray}
\nonumber
\Omega_\mathrm{GW,KD} & =& \Omega^{\mathrm{st}}_{\rm GW} h^2 (f_\Delta) \left(\frac{f_\mathrm{KD}}{f_\Delta}\right)   
 =  \Omega^{\mathrm{st}}_{\rm GW} h^2 (f_\Delta) e^{2 N_\mathrm{KD}}\\
 \nonumber
 &\simeq & 2.84 \times 10^{-13}  \left(\frac{g_*(T_\Delta)}{106.75}\right) \left(\frac{g_{*,s}(T_
\Delta)}{106.75}\right)^{-4/3} \nonumber \\
&\times&\left(\frac{V_{\inf}^{1/4}}{10^{16}\textrm{ GeV}}\right)^4 \left( \frac{\exp(2 N_\mathrm{KD})}{22000} \right),
\end{eqnarray}
where $\exp(10) \approx 22000$.
Finally:
\begin{align}
f_M  =  \frac{H_M a_M}{2 \pi a_0}  
=  f_\Delta  \left(\frac{\rho_M} {\rho_\textrm{KD}}\right)^{1/6}.
\end{align}
%
The amplitude difference between flat parts is
\begin{equation}
\frac{\Omega_\mathrm{GW}(f > f_M)}{\Omega_\mathrm{GW} (f<f_\Delta) }  =  \left( \frac{f_\mathrm{KD}}{f_\Delta}\right) \left( \frac{f_\mathrm{KD}}{f_M}\right)^2
=  \left( \frac{1}{\rho_\Delta} \cdot \frac {\rho_\textrm{KD}^2} {\rho_M}\right)^{1/3}.
\end{equation}
%
\begin{figure}[t]
\includegraphics[width=0.485\textwidth]{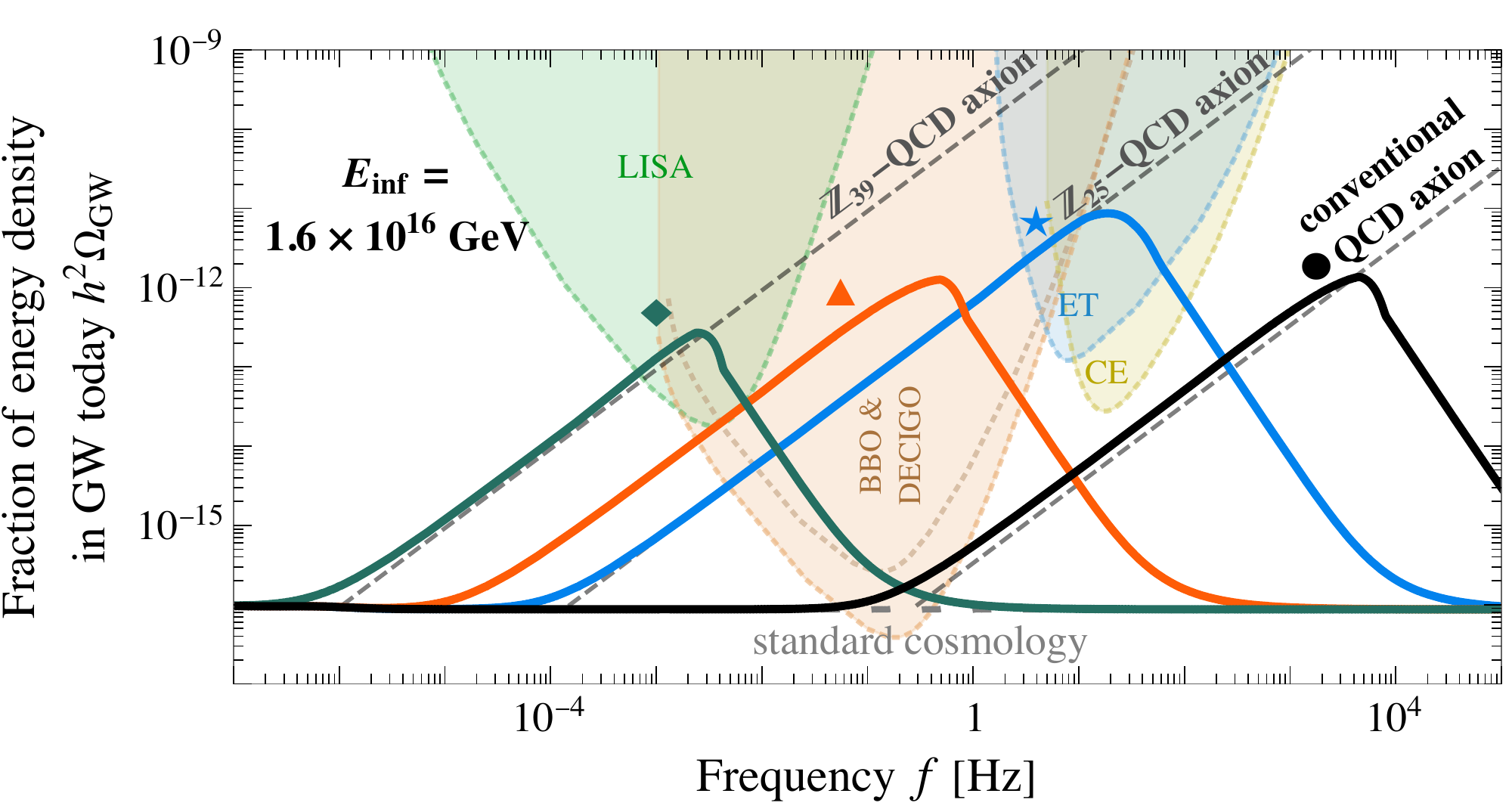}
\caption{\small \it{
The matter-kination scenario leads to a peaked GW spectrum from primordial inflation. The peak's height and position are determined by the inflationary scale $E_\mathrm{inf}$, the kination e-folding $N_\mathrm{KD}$, and the kination energy scale $E_\mathrm{KD}$. The shown spectra correspond to the benchmark points in Fig.~\ref{fig:detection}. The dashed lines represent the positions of the peak generated in different models of QCD axion dark matter (according to Eq.~\ref{eq_peak_position_ALP}).}}
\label{fig:spectra}
\end{figure}
If no entropy dilution occurs after the matter domination era, $\rho_\Delta = \rho_\mathrm{KD}^2/ \rho_M$ and the above ratio equals to unity.
The resulting typical spectra are plotted in Fig.~\ref{fig:spectra} for three benchmark points reported in 
Fig.~\ref{fig:detection} and corresponding to different choices of kination energy scales and kination duration. The shape of this spectrum is quite unique and very different from any other predictions of stochastic GW signals of cosmological origin. For instance, the peak that results from a cosmological first-order phase transition is very narrow as the source is active at a specific temperature  \cite{Caprini:2019egz} while here the kination era responsible for the peak lasts for several efolds. Another main source is cosmic strings. In this case, the source is long-lasting. The spectrum has a very different shape. It may feature a peak-like structure \cite{Gouttenoire:2019kij}, depending on the precise cosmological history. It can be close to scale-invariant over some frequency range, while the slope at low-frequency slope rises as $\propto f^{3/2}$. The effect of a kination era on the GW spectrum from cosmic strings are presented in a sister publication \cite{secondpaper}. If such cosmic string source is present, a multiple-peak structure may arise. 
Finally, another source of stochastic GW may come from the couplings of the inflaton. A well-known example is axion inflation that may lead to an enhanced signal due to parametric resonance effects induced by the inflaton coupling to gauge fields \cite{Barnaby:2010vf}. The spectral shape of this signal is also very different from what we predict from a short kination era.
In this letter, we focus on the model-independent irreducible background from inflation.
Fig.~\ref{fig:detection} shows which types of cosmological histories, characterised by the energy scale of kination and duration of kination, can be probed by LISA \cite{Audley:2017drz}, BBO \cite{Yagi:2011wg}, ET \cite{Hild:2010id, Punturo:2010zz}, CE \cite{Evans:2016mbw}  and SKA \cite{Janssen:2014dka}. To derive these regions, we have used the integrated power-law sensitivity curves of \cite{Gouttenoire:2019kij}.
Note that a kination era lasting more than $\sim 12$ efolds is not viable as it would lead to a too large energy density  in GW, violating theextra relativistic-species ($N_{\rm eff}$) constraint from Big Bang Nucleosynthesis (BBN) \cite{Caprini:2018mtu}.

Having derived the GW smoking gun signature resulting from an intermediate matter era followed by the kination era inside the radiation era, we will next argue that such cosmological history is a characteristic feature of axion field dynamics, that arises for instance in the Peccei-Quinn framework
before the axion starts oscillating and relaxes the strong CP parameter to unobservably small values.
Our discussion is very general and applies to any axion-like particle (ALP), the PQ axion being one particular example.
We will discuss two possible implementations. 
The first implementation relies on the interplayed dynamics of the radial and angular modes of the PQ field. A large kinetic energy can be transferred to the axion by the dynamics of the radial mode at early times.
The second one called ``trapped misalignment" only involves the axion, the angular mode of the complex PQ field, and was introduced in Ref.~\cite{DiLuzio:2021pxd,DiLuzio:2021gos}. In this framework, the axion has a large mass $M_a$  at early times. At some temperature $T_c$, the axion potential vanishes abruptly. In this process, the axion acquires a large kinetic energy that induces a kination era.
In both cases, the kination era lasts a few efolds, until the energy density of the scalar field, which redshifts as $a^{-6}$ is overcome by radiation, and the standard evolution then takes over.
Before  describing these two cases in turn, we investigate the general case where the axion makes all the dark matter (DM) of  the universe, as this represents one of the golden scenario that has led to a large variety of experimental searches.

\begin{figure}[t]
\begin{center}
\includegraphics[width=0.475\textwidth]{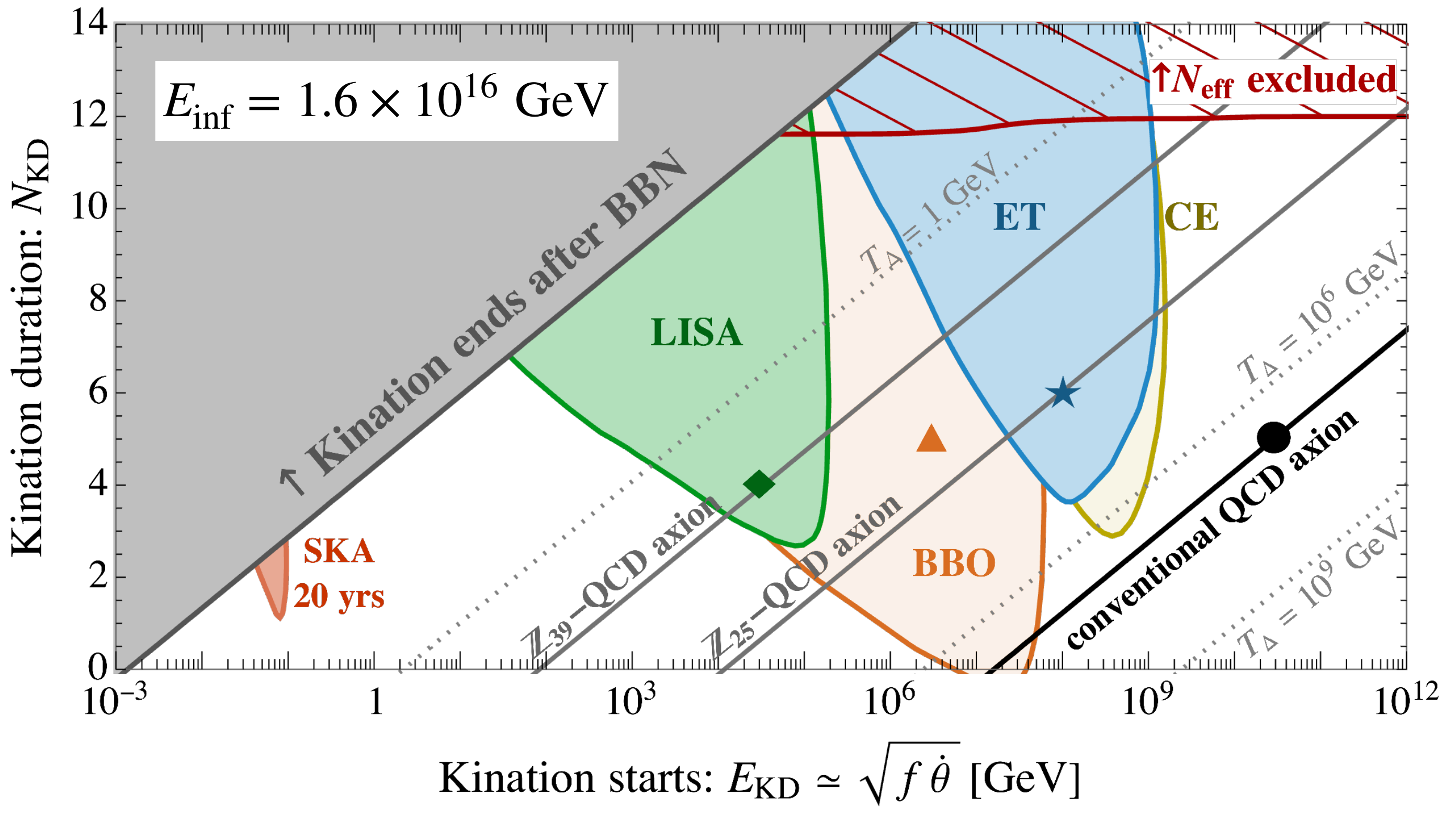}
\end{center}
\caption{\small \it{Model-independent probes of a short kination era in the early universe by GW experiments. Coloured regions indicate observable windows for each experiment. BBN constrains the energy scale at which kination ends (gray) and the amount of GW (red-hatched). 
Dashed lines indicate the temperature $T_{\Delta}$ when kination ends. Peaked signals exist in the white region but are not observable in planned experiments. Like in Fig.~\ref{fig:spectra}, this figure does not assume anything about axions, it just relies on a kination era as defined in Fig.~\ref{fig:diag}. Only the three parallel solid lines refer to specific models where the kination era is triggered by a QCD axion. The black line denotes the scenario where kination is induced by the spinning of conventional QCD axion DM, the corresponding GW peaks would require new observatories sensitive to ultra-high frequencies.
The  lighter QCD axion DM with $\mathbb{Z}_\mathcal{N}$-symmetry \cite{DiLuzio:2021pxd, DiLuzio:2021gos} can induce a GW signal, from the shown benchmark points, e.g.  at ET, BBO, and LISA for $\mathcal{N} \simeq 25, ~ 31, ~ 39$, respectively.}}
\label{fig:detection}
\end{figure}

\section{Gravitational waves and  axion Dark Matter relation}

\begin{figure}[t]
\begin{center}
\includegraphics[width=0.45\textwidth]{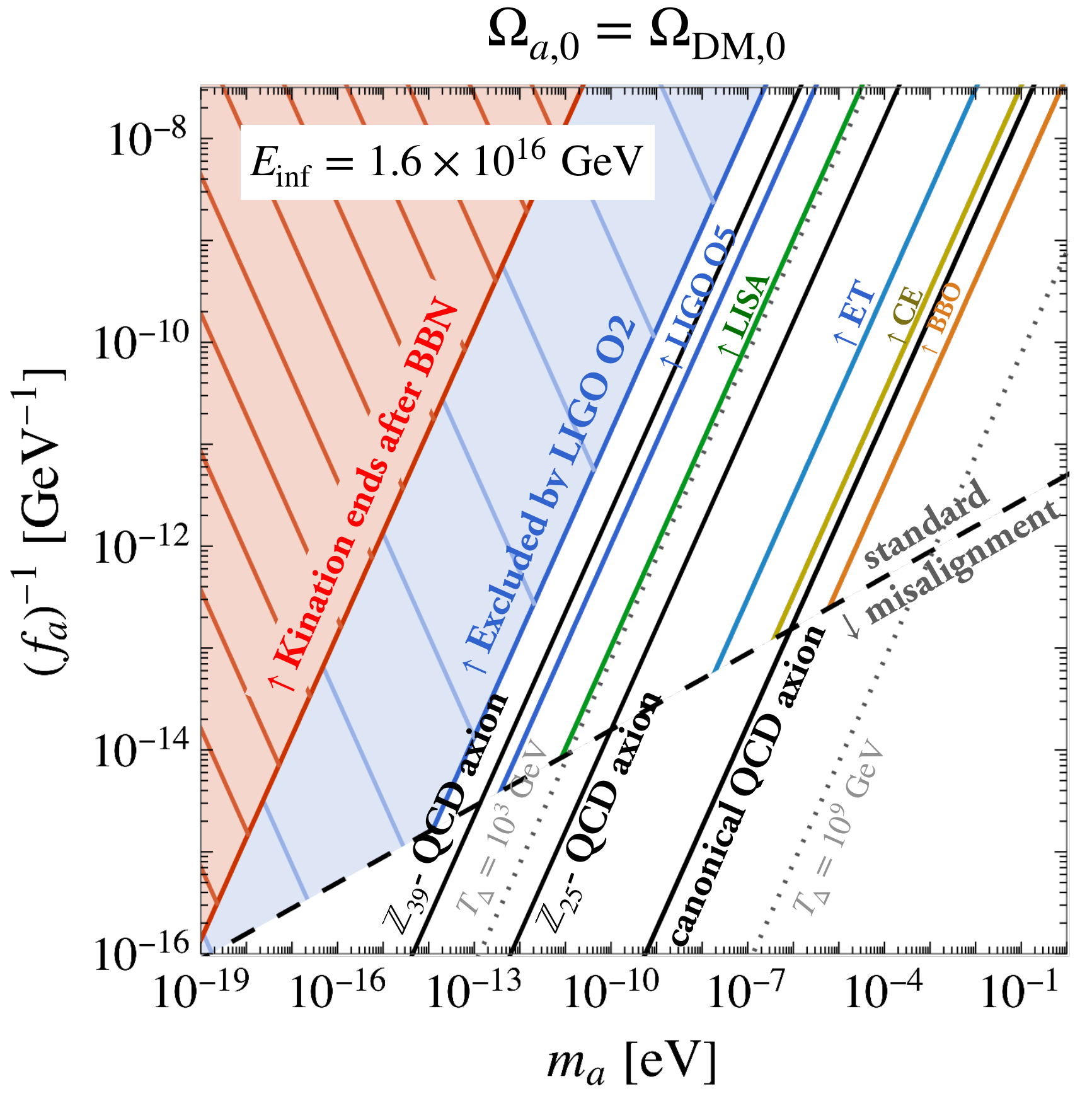}
\end{center}
\caption{\small \it{Ability of future GW experiments to probe a generic spinning ALP-DM as well as QCD axion-DM, whose abundance is set by kinetic misalignment. The peaked GW is observable on the left side of each of the  colored lines. Hatched regions are excluded experimentally (LIGO O2 \cite{Aasi:2014mqd}) and theoretically (kination era after BBN). The BBN bound includes the case where axion is trapped before kination ends. For larger $f_a$ or $m_a$, the kinetic misalignment is not effective (below gray-dotted line). The QCD axion mass relations are shown in black lines for the canonical (solid) and the $\mathbb{Z}_\mathcal{N}$-axion models (dashed).}}
\label{fig:axion_constraint}
\end{figure}

From the so-called kinetic misalignment mechanism \cite{Co:2019jts, Chang:2019tvx} or kinetic axion fragmentation  \cite{DESYfriendpaper}
the ALP relic abundance today  reads
 ${\Omega_{a,0}}/{\Omega_{DM,0}}  \simeq  170.94\left({m_a}/{1  \mathrm{eV}}\right) \left({Y_a}/{40}\right)
 $,
where the comoving axion number density is conserved after kination has started,  
$Y_{a}= n_{a}/s= f_a^2 \dot{\theta}_{\mathrm{KD}}/s(T_{\mathrm{KD}})$.
Eq.~(\ref{bump_peak_kination_inf}) can be re-written in terms of $Y_{a}$,
%
$
f_\mathrm{KD} = 4.6 \times 10^{-9}  \mathrm{Hz}  G^{1/4}(T_\Delta) G^{3/4}(T_\mathrm{KD}) \left({f_a}/{Y_a } \right) e^{2 N_\mathrm{KD}},
$
%
such that we can relate the GW peak amplitude to the ALP abundance today:
\begin{eqnarray}
\Omega_\mathrm{GW,KD} h^2  =&  (6.48 \times 10^{-19}) \left( \frac{G(T_\Delta)}{G(T_\mathrm{KD})} \right)^{3/4} \left( \frac{E_\mathrm{inf}}{10^{16} ~ \mathrm{GeV}} \right)^4 
\nonumber
\\
 \times & \left( \frac{f_\mathrm{KD}}{1 ~ \mathrm{Hz}} \right)  \left( \frac{10^{9} ~ \mathrm{GeV}}{f_a} \right) \left(\frac{1 ~ \mathrm{eV}}{{m_a}}\right) \left(\frac{\Omega_{a,0}}{\Omega_{DM,0}} \right).
\label{eq_peak_position_ALP}
\end{eqnarray}
From this, we can deduce a bound on the axion mass for which a given GW experiment will be able to probe the GW peak from a kination era induced by the axion of a given relic abundance. We report these bounds in the case where the axion  accounts for all the dark matter in the universe in Fig.~\ref{fig:axion_constraint}.

\section{Kination from a rotating complex scalar field }

We consider a complex scalar field $\Phi$ with a Lagrangian 
\begin{equation}
\mathcal{L}  =  (\partial_\mu \Phi)^\dagger \partial^\mu \Phi - V(\left| \Phi \right|) - V_{\cancel{U(1)}}(\Phi),
\end{equation}
where $V$ is a globally $U(1)$-symmetric potential with spontaneous-symmetry breaking (SSB) vacuum, and $ V_{\cancel{U(1)}}$ is an explicit $U(1)$-breaking term. 
The complex scalar field can be written as two real fields describing the radial $\phi$ and angular $\theta$ directions
\begin{equation}
\Phi =  \phi e^{i \theta},
\label{pfield_parametrize}
\end{equation}
where the $U(1)$-symmetry acts as a shift symmetry of  $\theta$.
In most of the literature on axion cosmology, the dynamics of the radial mode is not considered. One focuses on the oscillations of the axion at late times, once the radial mode has settled to its present value $f_a$.
On the other hand, the early dynamics of the axion's companion, the radial mode $\phi$, can be of crucial importance to 
motivate the initial conditions for the axion oscillations. As  first pointed out in \cite{Co:2019jts,Chang:2019tvx} and later exploited in \cite{Co:2020dya,Co:2020jtv}, the axion can indeed acquire a large initial velocity due to the early dynamics of the radial mode of the Peccei-Quinn field. 
In this section, we strongly rely on such general framework and show how a kination era is naturally induced.
In this case, the PQ symmetry is spontaneously broken before or during inflation and $\phi$ acquires a large VEV 
during inflation due to a Hubble-size negative mass induced by Planck-suppressed operators \cite{Affleck:1984fy, Dine:1995uk, Dine:1995kz}. From this large initial VEV, it  starts its evolution towards the minimum of the potential at $\phi=f_a$. Once it reaches the bottom, its kinetic energy dominates over the potential energy and allows the period of kination.
\begin{figure}[t]
\includegraphics[width=0.3\textwidth]{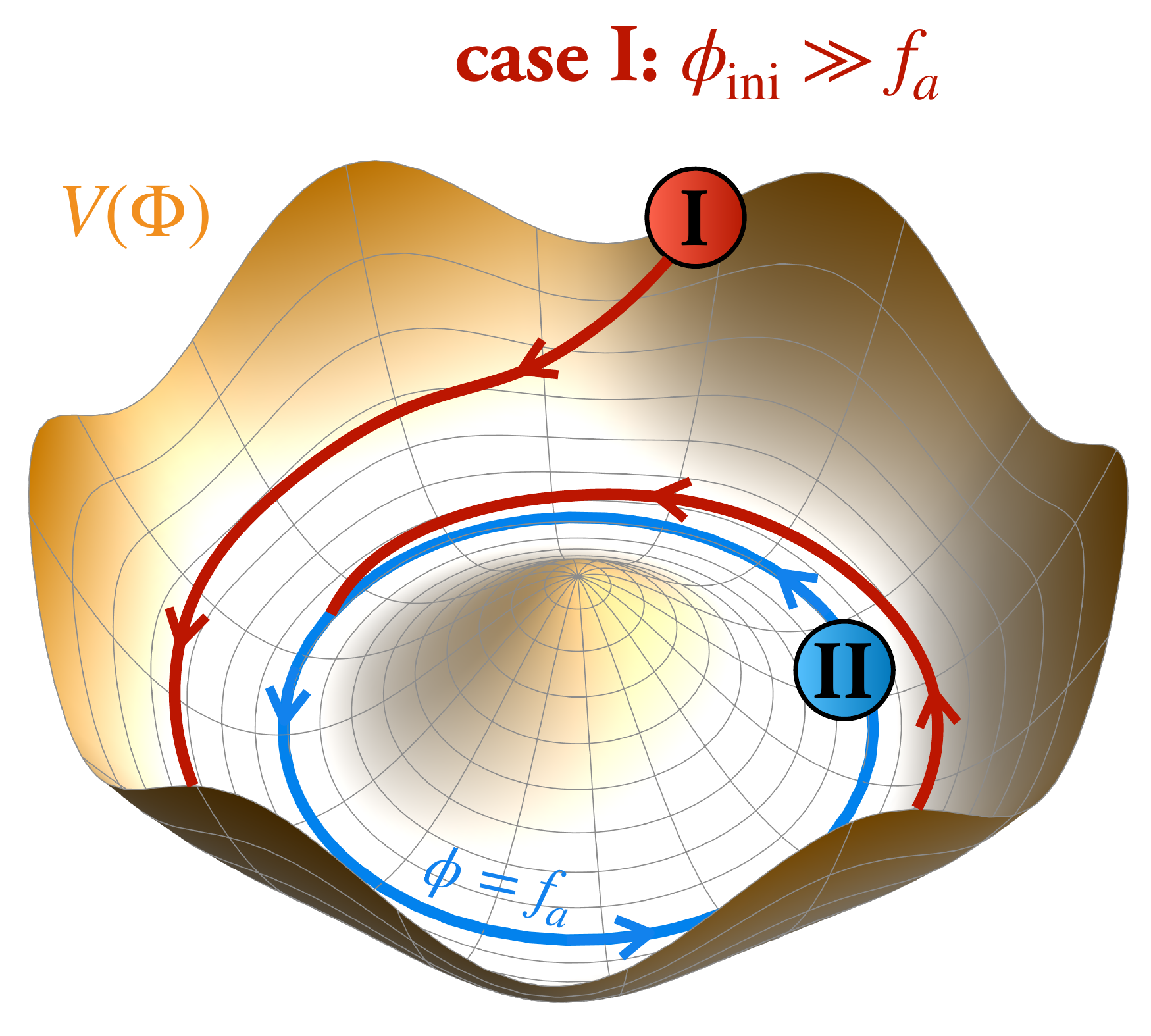}\\
\includegraphics[width=0.4\textwidth]{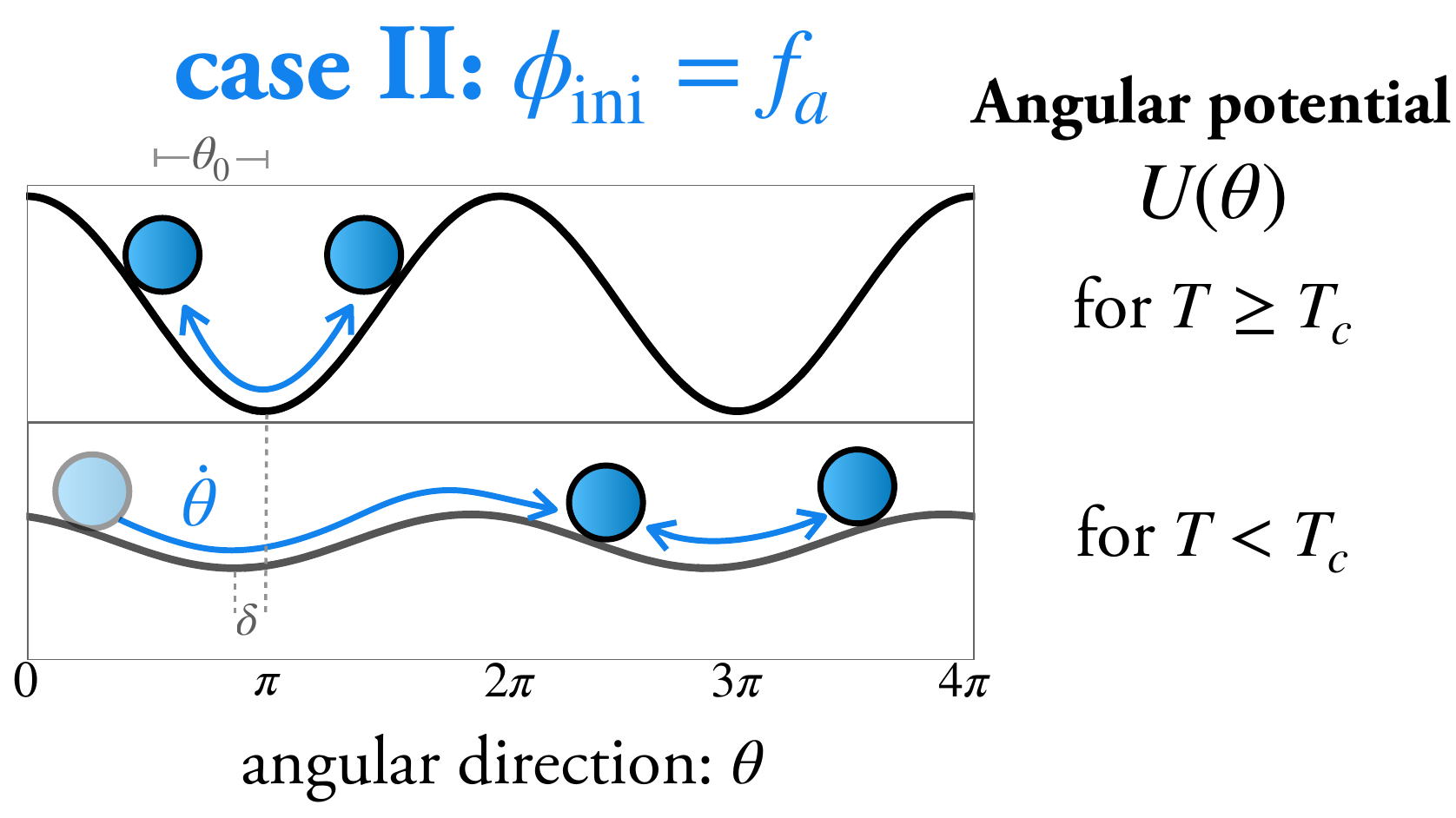}
\caption{\small \it{
Field-space evolution generating a kination era in two classes of models: (I) rotating complex condensate, relying on the interplayed dynamics between the radial and angular modes of the PQ field and (II) trapped misalignment, that only involves the angular mode dynamics, controlled by an abrupt change in the axion mass at some temperature $T_c$.
}}
\label{fig:trajectory}
\end{figure}
%
The kinetic energy induced by $ V_{\cancel{U(1)}}$ is stored by the field motion in this flat direction, which is crucial for kination. 
We consider only the homogeneous part of the field, the Lagrangian in the angular representation is 
\begin{equation}
\mathcal{L} =  \frac{1}{2}\dot{\phi}^2 + \frac{1}{2}\phi^2 \dot{\theta}^2  - V(\left| \Phi \right|) - V_{\cancel{U(1)}}(\Phi),
\label{lagragian}
\end{equation}
where the first and second terms denote the kinetic energy in the radial and angular modes, respectively. 
%
In the absence of the explicit breaking,  the angular equation of motion can be written as a constant of motion
\begin{equation}
\frac{d}{dt}\left(a^3 \phi^2 \dot{\theta}\right) ~ =~ 0,
\label{PQ_charge_conservation}
\end{equation}
where $\phi^2 \dot{\theta}$ is  the comoving conserved charge corresponding to the restored $U(1)$-symmetry.

In this story, the kination era occurs when the rotating field that dominates the universe settles down to the SSB minimum. The rotation of the axion field is generated similarly to the Affleck-Dine  mechanism \cite{Affleck:1984fy} where the explicit breaking potential imparts a kick in the angular direction. 
In contrast, the $U(1)$-symmetric potential excites the radial motion. With kicks in both directions, the field behaves as a coherent condensate with elliptic motion. For a nearly-quadratic potential, the radial oscillation induces an equation of state close to $\omega=0$. If this matter stage lasts long enough, the scalar field energy density can dominate the total energy density of the universe. 
We show the field trajectory in Fig.~\ref{fig:trajectory}-top.
A damping process is needed to suppress the oscillation along the radial direction $\dot{\phi} \to 0$. This can happen from interactions between the condensate and the thermal bath or from parametric resonance effects. 
These effects are studied in \cite{secondpaper}. Under these circumstances, the field accomplishes a circular orbit, where centrifugal force and curvature of the potential balance each others, and whose size decreases with time due to Hubble friction
\begin{equation}
\dot{\theta} = \sqrt{V^{''}(\phi)} \qquad \textrm{and} \qquad \phi \propto a^{-3/2}.
\end{equation}
When the circular orbit reaches the bottom of the potential $\phi \to f_a$, the energy density of the universe becomes dominated by the kinetic energy of the angular field, and starts evolving with a kination equation-of-state
\begin{equation}
\dot{\theta} = a^{-3} \qquad \textrm{and} \qquad \rho \propto a^{-6}.
\end{equation}
In summary, the following conditions must be satisfied:
\begin{itemize}
\item a $U(1)$-conserving potential  $V$ with spontaneous breaking which is nearly-quadratic in order to generate a matter era,
\item an explicit $U(1)$-breaking potential $V_{\cancel{U(1)}}$ to induce the angular motion at early time,
\item a large initial radial field-value $\phi_\textrm{ini} \gg f_a$, and a small angular displacement $\theta_\textrm{ini} \neq 0$, for $V(\phi_\textrm{ini})$ to be large and $V_{\cancel{U(1)}}(\phi_\textrm{ini})$ to be non negligible,
\item a damping mechanism for the radial mode only in order to circularize the trajectory.
\end{itemize}
The scalar field at $\Phi_\mathrm{ini}$ starts to move when the Hubble friction becomes smaller than the potential curvature along the radial direction, i.e., around the temperature 
$T_\mathrm{kick} ~ \sim ~ \sqrt{m_r M_{\rm pl}}$, with
\begin{equation}
m_r(\phi) \equiv \sqrt{V^{''}(\phi)}.
\end{equation}
At the same time, the field gets kicked along the angular direction by the $U(1)$-breaking term.
If $V_{\cancel{U(1)}}(\Phi_{\rm ini})$ and $V(\phi_{\rm ini})$ are comparable, then the initial kick speed $\dot{\theta}$ is of order $m_{r}(\phi_{\rm ini})$. In this letter, we consider this case for simplicity. A smaller ratio suppresses the rotational speed and, hence, the following kination energy scale and duration.  We leave further discussions to the companion paper \cite{secondpaper}.
After a few Hubble times of evolution, the explicit breaking term of order $(\phi/M_{\rm pl})^l$ for $l>4$ decouples from the equations of motion  and the $U(1)$ charge conservation law in Eq.~\eqref{PQ_charge_conservation} is restored.

The complex scalar field dominates the energy density of the universe and generates a matter-domination era at the energy density 
\begin{equation}
\rho_M~ \simeq ~ m_r^2 \phi_\mathrm{ini}^2 \left(\frac{\phi_\mathrm{ini}}{M_{\rm pl}}\right)^6.
\end{equation}
The matter era transits into a kination stage when the circularly-rotating complex scalar field reaches the flat direction of the potential $\phi \rightarrow f_a$, at the energy density
\begin{equation}
\rho_\mathrm{KD}~ \simeq ~ m_r^2 f_a^2.
\end{equation}
The kination era ends when the radiation energy density takes over. For simplicity, we neglect the entropy injected into the radiation bath during radial damping, which is the case if damping occurs before domination. So the radiation energy density scales as $a^{-4}$ from the time the scalar starts rolling to the kination ending. 
The energy density at the end of the kination stage is
\begin{equation}
\rho_\Delta~ = ~ \rho_\mathrm{KD}^2/\rho_M ~ \sim ~ m_r^2 f_a^2 \left(\frac{f_a}{\phi_\mathrm{ini}}\right)^2 \left(\frac{M_{\rm pl}}{\phi_\mathrm{ini}}\right)^6,
\end{equation}
and  the number of e-folding of kination era reads
\begin{equation}
\exp(N_\mathrm{KD}) ~ = ~ (\rho_M/\rho_\mathrm{KD})^{1/6} ~ \sim ~ \left(\frac{\phi_\mathrm{ini}}{f_a}\right)^{1/3} \left(\frac{\phi_\mathrm{ini}}{M_{\rm pl}}\right).
\end{equation}
In order to allow for the presence of a kination era, the initial field value should satisfy $\phi_\mathrm{ini} \gtrsim M_{\rm pl}^{3/4} f_a^{1/4}$. 
We will give more details on the expected value for $\phi_\mathrm{ini} $ in \cite{secondpaper}.

\begin{figure}[h!]
\includegraphics[width=0.42\textwidth]{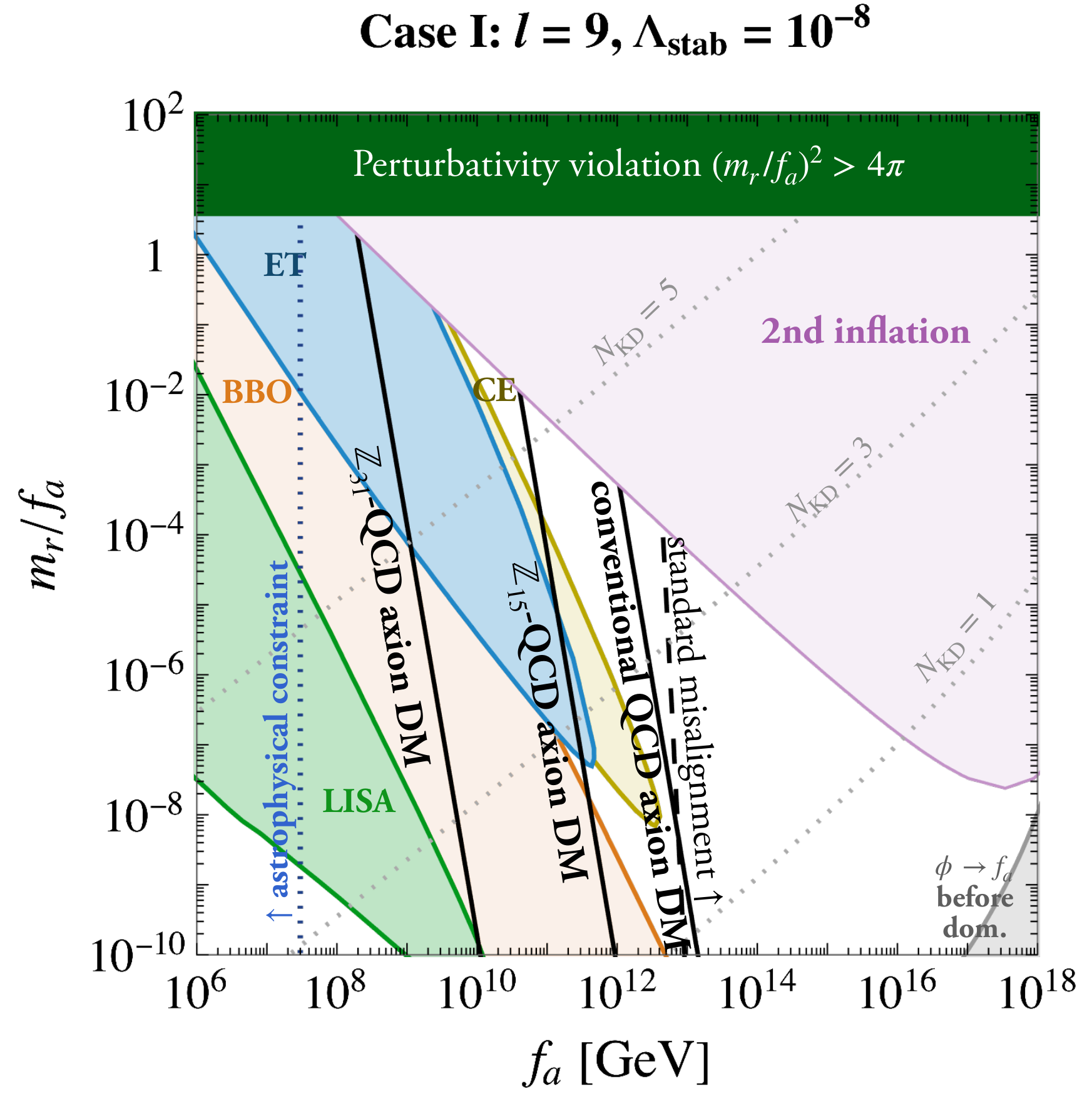}\\
\includegraphics[width=0.42\textwidth]{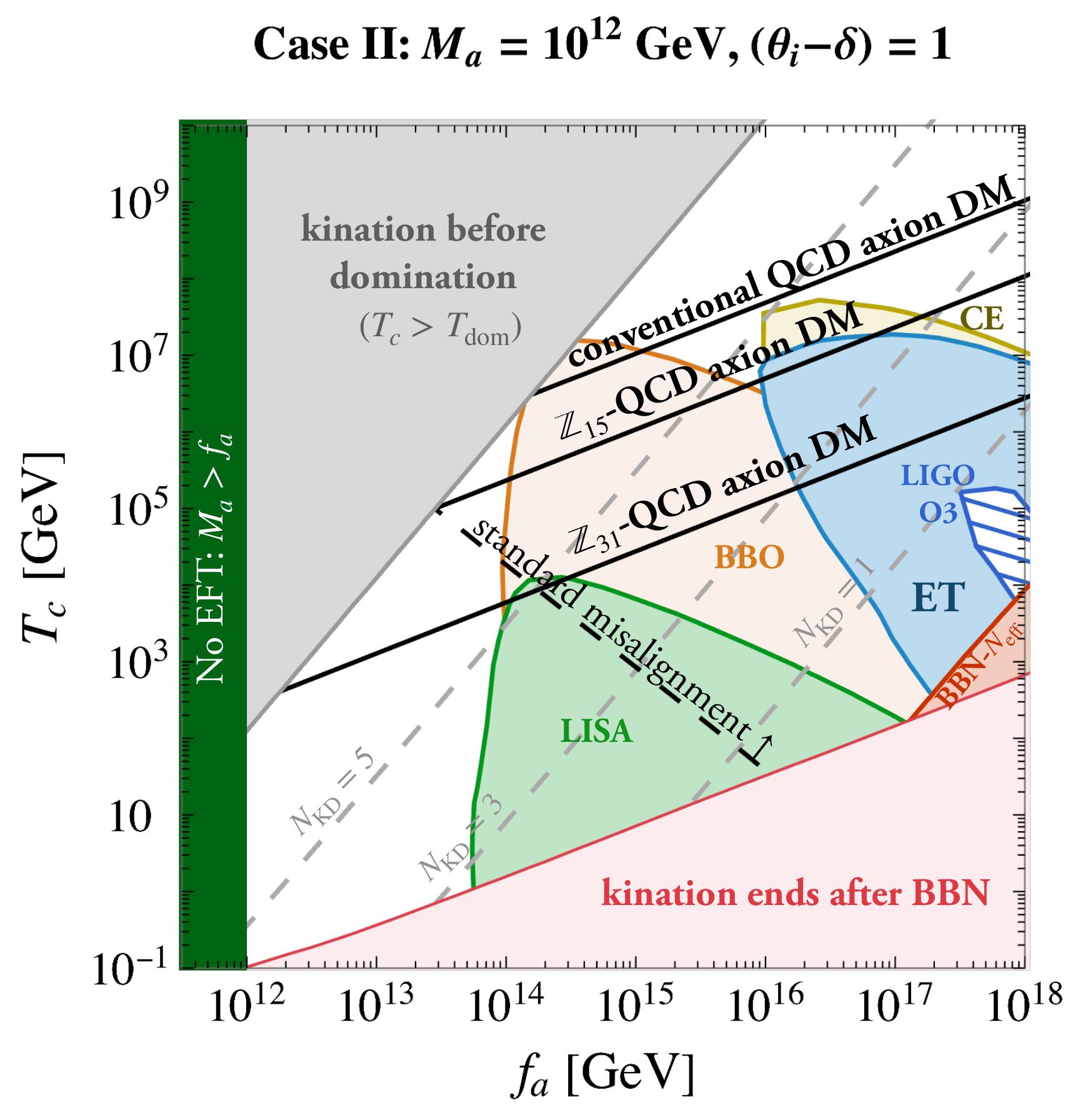}
\caption{\small \it{
The kination GW-peak allows  future observatories to probe the model of the rotating axion (top) and the trapped misalignment (bottom). 
Some regions of parameter space cannot consistently generate kination era, e.g., by violating BBN bound, generating second period of inflation, or violating EFT bound. The black solid lines denote the conventional and the $\mathbb{Z}_\mathcal{N}$-extension of the QCD axion. On the right of the black dashed line, the DM abundance is set by the standard misalignment mechanism. In case I, $ \Lambda_{\rm stab}$ parametrizes the size of  higher-dimensional (Planck-suppressed) stabilising operators in the scalar potential.
}}
\label{fig:model_dependent}
\end{figure}

In  Fig.~\ref{fig:model_dependent}-top, we show the region of parameter space in $(f_a,m_r/f_a)$ plane where the peak of the inflationary GW signal induced by the kination era is observable by either SKA, BBO, ET, CE and LISA.
We also indicate the contours where the QCD axion can lead to the correct DM abundance as discussed earlier.

%

\section{Kination from trapped axion misalignment }

There is an alternative way to induce a kination era, that does not rely on the radial mode dynamics but only involves the axion \cite{DiLuzio:2021pxd,DiLuzio:2021gos}. In this case, we do not need to assume that the symmetry was broken during inflation. We can just consider the dynamics of the axion alone once the radial mode has already reached its VEV $f_a$. The extra assumption is that the axion mass acquires a large mass $M_a$ at early times  and therefore starts oscillating well before the QCD scale.   The scalar field is initially frozen with the energy density
$\rho_\mathrm{osc}  = U(\theta_i)\approx  \frac{1}{2} M_a^2 f_a^2 (\theta_i - \delta)^2,$
where $\delta$ denotes the shift between the early-time and late-time minima and where we work in the small-misalignment limit.
The field starts moving when the Hubble rate drops to $3H \sim M_a$, or equivalently when the energy density of the background radiation is
$\rho_\mathrm{osc}^\mathrm{rad}  =  \frac{1}{3} M_{\rm pl}^2 M_a^2$.
%
As the field oscillates in the effective quadratic potential and redshifts as pressure-less matter, it eventually dominates the Universe.  At some lower temperature $T_c$, the cosine potential vanishes. The field then moves freely and drives a kination era if the kinetic energy exceeds the vacuum potential, see Fig.~\ref{fig:trajectory}-bottom. The energy density of the field at $T_c$ is 
$\rho_{\mathrm{KD}}  = \rho_\mathrm{osc}  \left(a_\mathrm{osc}/a_c\right)^3,$
where $a_c$ is the scale factor when the thermal-bath temperature drops to $T_c$. 
On average over many oscillations, the kinetic energy of the field is half of the total energy, and the axion obtains a speed $\dot{\theta}_c \simeq \sqrt{2 \rho_{\mathrm{KD}}/f_a^2}$ when the cosine potential vanishes. The kination era starts with the energy density
\begin{equation}
\rho_{\mathrm{KD}} = \frac{1}{2}  (\theta_i - \delta)^2 f_a^2 T_c^2  \left(\frac{T_c}{M_{\rm pl}}\right) \left(\frac{M_a}{M_{\rm pl}}\right)^{1/2}   \left[\frac{\pi^2}{10} g_*(T_c)\right]^{3/4}. 
\end{equation}
Kination era ends when the radiation becomes dominant again. This happens when the e-folding of kination satisfies $\exp(N_\mathrm{KD}) ~ \simeq ~ \sqrt{\rho_\mathrm{KD} / \rho_{\mathrm{rad}}(T_c)}$,
\begin{equation}
\exp(N_\mathrm{KD}) ~ \simeq ~ (\theta_i - \delta)^2 \frac{f_a^2}{T_c}  \left(\frac{T_c}{M_{\rm pl}}\right) \left(\frac{M_a}{M_{\rm pl}}\right)^{1/2}   g^{-1/4}_*(T_c). 
\end{equation}
 Fig.~\ref{fig:model_dependent}-bottom  shows the GW observability regions in the $(f_a,T_c)$ plane for $M_a=10^{12}$ GeV.
We also indicate the contours where the QCD axion can lead to the correct DM abundance  either from   kinetic or standard misalignment mechanism~\cite{DESYfriendpaper}.
There is potentially a domain wall problem or constraints from isocurvature perturbations, which can be addressed in various ways~\cite{secondpaper}.

%

%
%

\section{Conclusion}

We showed that a short kination era in the cosmological history generates a substantial enhancement of the irreducible stochastic gravitational wave background from inflation, with a characteristic peaked spectrum that can be observed at the next generation of GW interferometers.
An intermediate kination era cannot be obtained by any random scalar field dynamics in the early universe. It requires some very specific scalar field evolution, which we have argued to be symptomatic of axion-like particles.  We illustrated the predictions for the generic ALP case as well as for the QCD axion. The observation of the peaked GW signal we have discussed above would be a unique signature of  ALP  dynamics in the early universe as model parameters can be extracted from precise measurement of the GW spectrum. The signature from Peccei-Quinn symmetry breaking on the other hand motivates the design of ultra-high frequency GW experiments \cite{Aggarwal:2020olq}.

\paragraph{Acknowledgements}
We thank Cem Er{\"o}ncel, Pablo Quilez, Ryosuke Sato, Philip S{\o}rensen for useful discussions.
This work is supported by the Deutsche Forschungsgemeinschaft under Germany Excellence Strategy - EXC 2121 ``Quantum Universe'' - 390833306.

\paragraph{Note added}
These results  have been public over the last months through slides presented, for instance, at GWMess2021, BSM2021, Invisibles2021, PASCOS2021, EPS-HEP-2021, SUSY2021.
While this paper was being finalised, Ref.~\cite{Co:2021lkc} appeared with some overlap.

\bibliography{biblioSPIN} 

\end{document}